\documentclass[letterpaper,dvipdfmx]{article}

\usepackage[margin=34mm]{geometry}

\usepackage{graphicx}
\usepackage{latexsym}

\usepackage[varg]{txfonts}
\makeatletter%
\input{ot1txtt.fd}
\makeatother%

\usepackage{url}

\newcommand{\urle}[1]{$\langle$\nobreak\url{#1}$\nobreak\rangle$}

\newcommand{\doi}[1]{DOI: #1}

\usepackage{listings}
\lstdefinelanguage{FSharp}{
  language=[Objective]Caml,
  morekeywords={else, if, in, let, fun, match, open, rec, then, int, string, unit, it, System, Object, Int32, Int64, String, Convert, List, Collections, ArrayList, Generic, Tuple, Math, BigMul, Max, ToInt32, IndexOf, Xml, Linq, XElement, Load, IO, Stream, TextReader, TextWriter, IEnumerable, Enumerable, Distinct, ElementAt},
  sensitive=true,
  morecomment=[l]{//},
  morecomment=[f][keywordstyle][2]{>},
  morestring=[b]{"},
  morestring=[b]{'}
}

\lstdefinelanguage{CSharp}{
  language=C++,
  morekeywords={else, if, for, while, do, using, public, private, protected, internal, static, abstract, sealed, class, struct, int, string, System, Object, Int32, Int64, String, Convert, List, Collections, ArrayList, Generic, Tuple, Math, BigMul, Max, ToInt32, IndexOf, Xml, Linq, XElement, Load, IO, Stream, TextReader, TextWriter, IEnumerable, Enumerable, Distinct, ElementAt},
  sensitive=true,
  morecomment=[l]{//},
  morecomment=[f][keywordstyle][2]{>},
  morestring=[b]{"},
  morestring=[b]{'}
}
 
\lstdefinelanguage{Favalon}{
  language=FSharp,
  morecomment=[l]{\#}
}

\usepackage{enumitem}
\setlist{topsep=3pt,parsep=0pt,partopsep=0pt,itemsep=2pt}
\setlist[itemize]{leftmargin=14pt,labelsep=6pt}
\setlist[enumerate]{leftmargin=15pt,label=\arabic*.,labelsep=4pt}
\setlist[description]{leftmargin=10pt,style=nextline}

\usepackage{syntax}

\usepackage{framed}
\usepackage{booktabs}

\newcommand{\textid}[1]{\relax{\ttfamily\slshape #1}\relax}
\newcommand{\textids}[1]{\textid{\fontsize{8.9pt}{3pt}\selectfont #1}\relax}

\usepackage{authblk}

\newcommand{\keyword}[1]{\vspace{8pt}\centerline{\small Keyword: #1}}
\newcommand{\tabref}[1]{\textbf{Table \ref{#1}}}
\newcommand{\figref}[1]{\textbf{Fig \ref{#1}}}
\newenvironment{acknowledgment}{\section*{Acknowledgment}}{}

\begin{document}

\title{\textbf{A Proposal for an Interactive Shell Based on\\a Typed Lambda Calculus}}

\author{Kouji Matsui}

\date{}

\maketitle

\renewcommand{\thefootnote}{}
\footnote[0]{Information Processing Society of Japan Special Interest Group on Programming, Chiyoda, Tokyo 101--0062, Japan}
\footnote[1]{The Open University of Japan, Chiba, Japan 261--8586, Japan}
\footnote[2]{It has been presented at Information Processing Society of Japan Programming Study Group-132nd Programming Study Group.}
\renewcommand{\thefootnote}{\arabic{footnote}}%

\begin{abstract}
  This paper presents Favalon, a functional programming language built on the premise of a lambda calculus for use as an interactive shell replacement. Favalon seamlessly integrates with typed versions of existing libraries and commands using type inference, flexible runtime type metadata, and the same techniques employed by shells to link commands together. Much of Favalon's syntax is customizable via user-defined functions, allowing it to be extended by anyone who is familiar with a command-line shell. Furthermore, Favalon's type inference engine can be separated from its runtime library and easily repurposed for other applications.
\end{abstract}

\keyword{functional languages, scripting languages, command line interfaces, type inference}

\section{Introduction}

Modern interactive shells (both POSIX~\cite{posixshell} and non-POSIX variants) are highly responsive to user input and provide extensive scripting capabilities. It is relatively easy for users to acquire shell scripts to perform operations that could not be achieved through a shell's built-in commands alone. Users can even combine shell scripts with external commands in a pipeline when faced with tasks that would be difficult to implement using those scripts alone.

However, shell scripts and external commands are linked together with file streams (serialized bytes of data) or by interpreting unstructured text data. This makes it impossible to use any metadata that may have originally existed; as a result, it is challenging to take actions based on file metadata or even to determine how data is intended to be used.

Functional programming languages that use a typed lambda calculus, on the other hand, apply types to the data that is passed between functions, allowing a programmer to examine that data's intended usage in advance. Furthermore, the compiler or interpreter can provide feedback to the programmer whenever there is a type mismatch. These kinds of language features are used by autocomplete and other functionality provided by code editors and integrated development environments (IDEs).

At the core of any functional programming language is the lambda calculus, which is written using simple expressions comprised of function definitions and function applications. If we were to imagine a language processor capable of handling both shell scripts and external commands uniformly as functions, we may see that an ordinary shell script's syntax can be reduced into a series of expressions in the lambda calculus (or that it is at least compatible with the lambda calculus). For example, consider the following shell script:

\begin{lstlisting}[language=bash,morekeywords={wc}]
  echo "abc def ghi" | wc
\end{lstlisting}

We can interpret this as a series of applications of predefined functions. Even if \textids{echo} was a built-in shell command and \textids{wc} was an external command, we could consider both to be functions and the vertical bar to be an operator function. In this way, an entire shell script can be interpreted as a collection of function expressions to which a typed lambda calculus can be applied, allowing the language processor to perform type inference, examine types, provide feedback to the user, and even safely insert additional processing that abstracts away complex operations based on byte arrays for particular data types.

In this research, we will see how existing shell script syntax can be applied as a lambda calculus to combine both built-in and external commands with associated types.

\section{Issues Related to Running Tasks in Shells}
\label{sec:workarounds}

Command-line shells are often used for routine tasks such as operating system management and general file manipulation. Each operating system includes its own standard shell interface that users can leverage to achieve their goals. If the standard shell interface is insufficient to accomplish a particularly complex task, however, users must resort to the following workarounds:

\begin{enumerate}
  \item Work with external commands that manipulate files and text. For example, the user could manipulate text data with an external command that handles text processing. These command invocations can be combined using pipes in the shell.
  \item Create external commands with some other compiler or interpreter in a programming language that provides high-level, feature-rich libraries. This approach may be taken if it is too cumbersome to combine the external commands described above or if those commands are too slow.
  \item Use extended shell programming features. A flexible feature set makes it much easier to accomplish a task.
  \item Extend the shell itself.
\end{enumerate}

The following sections identify the pain points that are mitigated by each of these approaches before summarizing the problems and other issues that they entail.

\subsection{Combining Shell Scripts With External Commands}

Command execution is a basic feature of command-line shells. Built-in commands such as \textids{echo} and \textids{set} are used to perform tasks, but there is a limit to what can be done with these commands alone; the user is expected to combine command invocations with shell script control structures to overcome these limitations.

Bash scripts and many other shell scripts include \textids{for}, \textids{if}, and other flow control statements to implement complex loops and conditional logic. However, it is difficult to write code that is dependent on---or is responsible for manipulating---file data. In these situations, more advanced data processing can be achieved through a combination of shell scripts and external commands (the latter of which are implemented in some other programming language).

The first workaround listed in \S\ref{sec:workarounds} allows text data to be modified by \textids{sed}, \textids{awk}, and other external commands provided by default on POSIX-compliant operating systems. Pipes and redirection facilities are also provided by the shell to allow these commands to be linked together.

The second workaround listed in \S\ref{sec:workarounds} involves implementing new external commands in programming languages that include libraries with more features, better performance, and higher-level APIs. Because command-line shells and external commands communicate with each other using byte streams through standard input and output channels, commands can be implemented in any programming language that supports standard input and output; available options include C, C++, Perl, Go, and Python, to name a few.

With libraries that can manipulate data more quickly and at a higher level of abstraction, these programming languages can be used to implement operations that would be difficult to accomplish with shell scripts. C and C++ can process massive amounts of binary data; Perl excels at manipulating text; Go offers a concise syntax for directly making system calls and accessing Internet resources; and Python is well-suited to numerical calculations and machine learning.

These programming languages are also capable of invoking other external commands. For example, the following snippet of Go code uses the \textids{goexec} package~\cite{goexec} to call an external command and store the result in a variable:

\begin{lstlisting}[language=C++]
  // Run ls -la
  out, err := exec.Command("ls", "-la").Output()
\end{lstlisting}

Though some programming languages compile code into executable files, others provide an interactive environment known as a read–eval–print loop (REPL) that accepts user commands as input, evaluates those commands in the corresponding programming language, and returns the results as direct feedback to the user. Examples of REPLs include \textids{gore}~\cite{gore} (for Go) and F\# Interactive~\cite{fsinteractive}.

Each REPL generally supports either a full programming language specification or a slightly smaller subset of the language. For example, F\# Interactive uses double semicolons (';;') instead of newline characters to terminate entries because F\# would ordinarily evaluate a line of code as soon as it encountered a newline character. Furthermore, when F\# Interactive encounters a pair of semicolons the REPL executes the input code and automatically binds the result to the \textids{it} identifier. The resulting syntax is concise and resembles command execution in an existing shell:

\begin{lstlisting}[language=FSharp]
  // This does not have to be explicitly bound to
  // a value with 'let', but the line must be
  // terminated with ';;'
  > System.Int32.Parse("123");;
  val it : int = 123
\end{lstlisting}

Python has an interactive web-based environment called Project Jupyter~\cite{jupyter} that is based on IPython~\cite{ipython}; it allows users to be highly productive and expressive while editing code in ``notebook'' documents. There is also a vast Python library ecosystem that can be leveraged in many different kinds of applications.

\subsection{Extended Shell Programming Features}

Shell aliases are one example of the third workaround listed in \S\ref{sec:workarounds}. Using the Bash \textids{alias} command makes it appear as if you have defined a brand new command; for example, the alias defined below lists results in long format for the path given as an argument:

\begin{lstlisting}[language=bash,morekeywords={alias}]
  # List files and directories in long format
  alias lsla='ls -la'
\end{lstlisting}

This is an example of simple command substitution, but we can also use the \textids{function} keyword to define a custom function with more advanced logic:

\filbreak

\begin{lstlisting}[language=bash,morekeywords={function}]
  # List files and directories in long format
  # 10 times
  function lsla10() {
    for i in {1..10} ; do
      ls -la $1
    done
  }
\end{lstlisting}

Bash provides flow control statements that, together with function definitions, allow us to define commands that perform complex operations.

PowerShell~\cite{monadpsh} is another example of a practical command shell; it uses the .NET runtime and supports a first-order generic type system. The PowerShell language syntax even allows the user to specify types, as shown here:

\begin{lstlisting}[language=bash,breaklines=true,morekeywords={GetTypes,&}]
  # List all the types defined by the .NET
  # standard library and then count the number
  # of lines
  PS C:\> [System.Type]::GetType("System.Object")
  .Assembly.GetTypes() | &
  'C:\msys64\usr\bin\wc.exe'
     3302   12871  511351
\end{lstlisting}

The \textids{GetTypes} method returns an array of type \textids{System.Type[]} that can be combined with later stages of the pipeline. The ampersand ('\&') then calls an external command and passes it a string through the standard input stream; in this case, PowerShell has automatically coerced the array returned by the \textids{GetTypes} method into a string. Though this example gives a result of 3302 lines, that isn't an accurate number because it includes the ``header'' and ``footer'' in the automatically converted text.

Compared to ordinary shells, the syntax here may be a bit clunky and the text conversion is not precise, but PowerShell does make it easy to combine .NET library methods with external commands in a type-safe way. PowerShell's ability to handle types is used along with type inference to interactively autocomplete types in the Windows PowerShell Integrated Scripting Environment (ISE)~\cite{psise}.

Shcaml~\cite{shcaml} is one example of prior work in programming languages that force the user to pay attention to types when invoking external commands. Shcaml augments OCaml with definitions for operator functions that resemble the operators seen in shell scripts and a \textids{Shtream} type that provides an abstraction for data pipelines. Consider the following script from Figure 2 on page 80 of \cite{shcaml}:

\begin{lstlisting}[language=FSharp,morekeywords={run,begin,end,-|,from_file,grep,cut,mail_to}]
  run begin
    from_file "/etc/passwd"                                -|
    Adaptor.Passwd.fitting ()                              -|
    grep (fun line -> Passwd.shell line = "/usr/bin/scsh") -|
    cut Passwd.name                                        -|
    mail_to ~subject:"A New Shell" ~msg:"Check out Caml-Shcaml!"
  end
\end{lstlisting}

This example script joins commands using the pipe operator defined in Shcaml as ('\verb#-|#'). Some of the commands are functions defined by OCaml and others are external commands. \tabref{tab:fitting} is an excerpt from Table 2 on page 84 of \cite{shcaml}, which shows some operators in Shcaml and their shell equivalents. The operators use slightly different symbols but are defined so that the functions on either side of them can be combined with matching types.

\begin{table*}[bt]
  \caption{Some of Shcaml's fitting combinators~\cite{shcaml}}
  \label{tab:fitting}
  \centering
  \begin{tabular}{lll}
    \toprule
    Fitting combinator & Usage & Shell equivalent\\
    \midrule
    \scriptsize \tt \verb!(-|)  : ('i -> 'm) t -> ('m -> 'o) t -> ('i -> 'o) t! & \scriptsize \texttt{$c_1$ -| $c_2$} & \scriptsize \texttt{$c_1$ | $c_2$} \\
    \scriptsize \tt \verb!(/>/) : ('i -> 'o elem) t -> dup_out_spec -> ('i -> 'o elem) t! & \scriptsize \texttt{$c$ />/[$d_1$\%>\&$d_2$]} & \scriptsize \texttt{$c$ $d_1$>\&$d_2$} \\
    \scriptsize \tt \verb!(^&)  : (text -> 'b elem) t -> ('i -> 'o) t -> ('i -> 'o) t! & \scriptsize \texttt{$c_1$ \^{}\& $c_2$} & \scriptsize \texttt{$c_1$ \& $c_2$} \\
  \bottomrule
\end{tabular}
\end{table*}

\textids{Shtream}s are defined as an OCaml algebraic datatype; the data format that they actually accept and pass along can be freely customized through type variables for \textids{channel} functions. These \textids{channel} functions can be thought of as an abstraction layer for the data in a stream as well as the method by which that data is passed in and out of the stream, thereby allowing  connections between functions to be flexibly redefined.

Because Shcaml is built on top of OCaml, two semicolons (';;') must be used to mark the end of an expression when it is used as a REPL just like F\# Interactive.

OCommand~\cite{ocommand} is another example of prior work in programming languages that strongly call the user's attention to types when invoking external commands. Of course, OCommand is not a language processor \textit{per se}---it is a domain-specific language (DSL) that uses OCaml to associate concrete types with the input and output of external commands and then track those types as the standard I/O streams and command-line arguments change. Consider the following sample code from Figure 1 on page 2 of \cite{ocommand} (comments added here for clarity):

\begin{lstlisting}[language=FSharp]
  (* Run ls -l and perform calculations based on
     file sizes in the result *)
  let collect_files size =
    let open Ls in
    (* Specify an option equivalent to -l *)
    let files = command (add l empty) in
    let rec iter cur_size acc files =
      match files with
      | [] -> acc
      | file :: files ->
        (* size can be safely accessed with any
           command-line arguments *)
        let cur_size = cur_size + file..size in
        if cur_size > size then acc
        else iter cur_size (file..name :: acc) files
    in
    iter 0 [] files
\end{lstlisting}

The command-line arguments to \textids{ls} determine its output format, so the DSL is used to identify the location of columns with information such as the file size printed by a call to \textids{ls -l} and make that information accessible as bound variables of the correct type.

The fourth workaround listed in \S\ref{sec:workarounds} involves modifying the shell itself to accomplish a task. Users may resort to this option if they are unable to successfully make use of the extended features described so far in this paper or if they find it difficult to do so. A canonical example of this is the development of tcsh~\cite{tcsh} (\textit{cf.} csh).

The tcsh shell provides an extended interactive feature set, including command-line completion, command history, and job control. tcsh shell scripts are compatible with csh syntax but also include several new built-in commands along with new read-only variables that reflect the current state of the shell.

\subsection{Problems and Issues}

The approaches discussed so far present the following problems and issues to consider:

\begin{itemize}
  \item It is not always possible to complete a task using the external commands provided by the operating system. For example, \textids{sed} and \textids{awk} offer a concise and flexible way to manipulate text data, but they are not suited to making more precise alterations or manipulating binary data. They are also rather slow.

  \item If users implement workarounds to the aforementioned problem in a programming language of their choice, they can no longer directly use the convenience features available in shell scripts. For example, shell scripts can use pipes and redirection to easily describe how external commands should be linked together, but these features are (presumably) not natively supported by the user's chosen programming language. As a result, more complex language-specific code must be written to do the same thing. Furthermore, because this code is maintained separately from shell scripts, its author must ensure that a compatible software version is being used.

  \item Users who wish to write code for a programming language with a REPL environment must not only learn a REPL-specific grammar for that language but also a different syntax than they may be accustomed to using in shell scripts.

  \item Python's Jupyter notebook environment uses a persistent, web-based document model that combines scripts with their output. It is unlike conventional command-line shells and thus cannot be used as a replacement.

  \item Because shell scripting languages generally don't include type information, shell scripts cannot provide feedback to the user or examine code using types. Some interactive shells do support command-line completion, but if type information is unavailable they can only list partial symbol name matches or use manual definitions~\cite{bashac}.

  \item PowerShell has built-in support for type information, but it is difficult to make full use of this information because of PowerShell's unique syntax and frequent need for explicitly-specified types.

  \item Because Shcaml is built on top of OCaml, it is very flexible, easy to extend, and able to trivially call other OCaml functions. Though its operators are similar to the ones used by POSIX shells, they cannot be used interchangeably. Furthermore, the same restrictions that apply to F\# Interactive also apply to the Shcaml REPL, which is implemented using OCaml.

  \item OCommand can associate types with external commands and perform type introspection, but it requires its users to manually create a DSL.

  \item Users can (theoretically) satisfy any feature request by altering the shell itself. However, they must balance their desire to do so against the difficulty involved in modifying the shell and the work involved in maintaining it afterwards. On top of that, they must continue to modify the shell as new extensions become necessary, causing it to inevitably diverge further and further from the original code base. Users of common interactive shells would ordinarily not go to this much trouble to accomplish a task.
\end{itemize}

\section{Purpose of This Research}

Given all the examples and sample code covered in the previous sections while also considering issues identified in prior work, we now consider Favalon, a proposal for an interactive shell and its language processing engine. Favalon

\begin{enumerate}
  \item is based on a typed lambda calculus and can emulate or extend ordinary shell syntax. The language itself does not provide a native implementation of shell syntax; instead, users can write their own extensions to produce syntax that approaches a variety of shell scripts for both POSIX and non-POSIX shells.

  \item can fully integrate with existing runtime libraries, which the language uses just like external commands.

  \item sequentially evaluates input and produces output as an interactive shell; it also supplies and suggests inferred types to the user.

  \item can evaluate input expressions step-by-step or generate a compiled script.

  \item \label{lst:purpose:5} is implemented as a core engine library that can easily be ported to other computing environments.
\end{enumerate}

The Favalon programming language is built on top of the .NET runtime. As a result, the aforementioned ``existing runtime libraries'' include all of the libraries and code available for .NET. By porting Favalon's core engine to other computing environments as described in item~\ref{lst:purpose:5}, users can easily add it to a variety of applications and thus expand the results of this research beyond the Favalon programming language alone.

A typed lambda calculus can express any operation through a combination of two types of calculations: function definitions and function applications. Function applications, in particular, can be written as an ordered list of bound functions. Shell script syntax, on the other hand, is mainly defined as a series of tokens delimited by whitespace. If we assume that each token represents a function and then treat the list of all tokens as function applications, we may be able to reduce a shell script's syntax into the lambda calculus.

If we can also assign the appropriate types to the function definitions, we can reap the following benefits:

\begin{itemize}
  \item We can apply type inference rules from a typed lambda calculus.

  \item We can choose the appropriate functions to bind by inference. This can apply equally to the selection of both external commands and the appropriate functions from runtime libraries by treating them all as functions.

  \item We can provide autocomplete features and other feedback to the user in the interactive shell.
\end{itemize}

In the process of drafting this proposal for the Favalon programming language, the author has written and tested C\# implementations of each of the algorithms described in the following sections.\footnote[3]{The source code is avaliable at \url{https://github.com/kekyo/Favalon/}}

\section{Structure of the Favalon Programming Language}

\subsection{Lambda Calculus Evaluation Engine}

Favalon is comprised of two components: a core lambda calculus evaluation engine (Favalet) and an application that serves as the frontend to the interactive shell (Favalon). The following sections simply use the name ``Favalon'' to collectively refer to both components.

Favalon is implemented as a library responsible for the following tasks:

\begin{enumerate}
  \item Lexical analysis: Generate an array of tokens from an input shell script based on an extremely simple lexical grammar.

  \item Parsing: Generate a syntax tree that considers almost every token to be a simple function application.

  \item Type inference and $\beta$-reduction: Perform full or partial $\beta$-reduction on the syntax tree to do type inference (including higher-kinded types).
\end{enumerate}

\subsection{Lexical Analysis}
\label{sec:lexer}

The lexical analyzer follows a set of rules to convert an input shell script into a sequence of tokens. \figref{fig:grammar} shows Favalon's lexical grammar, which was kept as simple as possible to make the language's syntax flexible and customizable.

\begin{figure*}[h]
  \centering
  \begin{minipage}{320pt}
    \begin{framed}
      \begin{grammar}
        <token> ::= [<identity> | <string> | <numeric> | <parenthesis> | <symbol>]+

        <identity> ::= <letter> [<letter> | <digit> | `\_']*

        <string> ::= `"' <text>* `"'

        <parenthesis> ::= <begin-parenthesis> | <end-parenthesis>

        <symbol> ::= [`!' | `@' | `\#' | `\$' | ... ]+
      \end{grammar}
      \vspace{-3pt}
    \end{framed}
  \end{minipage}
  \vspace{3pt}  
  \caption{Favalon's lexical grammar}
  \label{fig:grammar}
\end{figure*}

\begin{enumerate}
  \item ID tokens are represented by any sequence of alphanumeric and underscore characters.

  \item String tokens are represented by text surrounded by double quotation marks.

  \item Numeric tokens are represented by any sequence of digits, which may be preceded by a sign symbol (plus or minus) and/or include a decimal point.

  \item Parenthesis tokens are represented by any matching pair of Unicode characters with the General Category values ``Open Punctuation''~\cite{unicodepunctuationopen} and ``Close Punctuation''~\cite{unicodepunctuationclose}.

  \item Symbol tokens are represented by any sequence of symbols.
\end{enumerate}

For example, consider an ordinary shell script like the following:

\begin{lstlisting}[language=bash,morekeywords={echo,wc,|}]
  echo "abc def ghi" | wc
\end{lstlisting}

This shell script is converted into the following sequence of tokens:

\begin{lstlisting}[language=CSharp,morekeywords={Identity,String,Symbol}]
  Identity(echo),
  String("abc def ghi"),
  Symbol(|),
  Identity(wc)
\end{lstlisting}

\textids{if}, \textids{for}, and other flow control statements in shell scripts can all be replaced with ID tokens.

The lexer draws distinctions between ID, parenthesis, and symbol tokens so that it can tokenize strings even if they are not surrounded by whitespace. The parser considers these tokens to be the same.

\subsection{Parsing}

The sequence of tokens obtained from the lexer are converted into a syntax tree as follows, where the nodes of the tree are called ``\textids{Expression}'' and in particular single term is called ``\textids{Term}''\footnote[4]{It omitted at the code suffix following examples.}:

\begin{enumerate}
  \item All tokens are left-associative. All combined nodes become \textids{ApplyExpression}s.
  \item ID, parenthesis, and symbol tokens become \textids{VariableTerm}s.
  \item String and numeric tokens become \textids{ConstantTerm}s.
\end{enumerate}

Note that the associativity of an ID may not match what it refers to. For example, the binding operator ('=') and the lambda expression arrow ('\verb|->|') are normally right-associative, but these differences are ignored while parsing.

If there are two or more tokens in a row, they are connected to the tree as child nodes of \textids{ApplyExpression}. For example, the parser generates the following tree for the shell script in \S\ref{sec:lexer}:

\filbreak

\begin{lstlisting}[language=CSharp,morekeywords={Apply,Constant}]
  // VariableTerm expressions are omitted below
  Apply(
    Apply(
      Apply(
        echo,
        Constant(
          "abc def ghi",
          System.String)),
      |),
    wc)
\end{lstlisting}

An implicit type annotation of \textids{System.String} has been applied to \textids{ConstantTerm} here.

\subsection{Type Inference and $\beta$-Reduction}
\label{sec:inference}

Type inference and $\beta$-reduction are usually done separately, but they are grouped together here because Favalon sometimes performs $\beta$-reduction in the middle of type inference. This is also the point at which Favalon transforms the syntax tree to be right-associative wherever it encounters a right-associative token before continuing evaluation, a process described in more detail later on.

Favalon uses the Hindley-Milner type inference algorithm~\cite{algorithmw}; it also supports higher-kinded types and recursively defines any higher-kinded expressions. The following sections describe some of the typical expressions that Favalon operates on.

\subsubsection{Lambda Expressions}

\textids{LambdaExpression} represents a lambda expression with a parameter and a body, both of which contain higher-kinded expressions. A lambda expression's type is inferred as follows:

\begin{enumerate}
  \item Infer the parameter type.
  \item Capture the parameter by binding it to a variable.
  \item Infer the body type.
  \item If a type annotation has been applied to the lambda expression itself, infer that expression's type.
  \item Unify all of the above. The parameter and body expressions are temporarily combined into a \textids{LambdaExpression} before being unified with the lambda expression's type annotation.
\end{enumerate}

\subsubsection{Function Applications}

\textids{ApplyExpression} represents a function application. This is left-associative for all tokens; it comprises a function and an argument. Although the lambda calculus only uses function applications on lambda expressions, Favalon's actual runtime system uses function applications on any expression that it can---not just \textids{LambdaExpression}s---because it needs to make external function calls to library code (as explained in a later section).

\begin{enumerate}
  \item If a type annotation has been applied to the function application itself, infer that expression's type.
  \item Infer the argument type.
  \item Infer the function type by combining the inferred results from steps 1 and 2 into a \textids{LambdaExpression}.
  \item Unify all of the above.
\end{enumerate}

\subsubsection{Function Overloading}
\label{sec:overloading}

C\# and F\# allow programmers to call type member methods defined by the .NET library (hereinafter simply called ``methods'') just like ordinary functions. The .NET type system also allows methods to be overloaded with different parameters but \textit{not} with different return values alone~\cite{ecma335overload}.

F\#'s .NET methods use the tuple form of passing arguments so they cannot be curried and do not support partial function application. However, because each of these methods has a return value of a fixed type, this means that they will not violate any overloading restrictions. With the exception of .NET methods, F\# also does not permit more than one function of the same name to be bound in a single module---in other words, F\# functions cannot be overloaded.

By requiring all functions to have one parameter and one return value (unary functions of arity 1), Favalon prevents complexity from creeping into its scripting syntax. Functions (methods) defined in external libraries with multiple parameters are replaced with curried versions so that they all become unary functions. For example, consider the following methods from the .NET Class Library:

\begin{lstlisting}[language=CSharp]
  // (1)
  System.Int32 System.Convert.ToInt32(
    System.String value, System.Int32 fromBase);
  // (2)
  System.Int32 System.Convert.ToInt32(
    System.String value);
\end{lstlisting}

In Favalon, these are defined as follows:

\begin{lstlisting}[language=FSharp]
  // (1)
  System.Convert.ToInt32:
    System.String -> System.Int32 -> System.Int32
  // (2)
  System.Convert.ToInt32:
    System.String -> System.Int32
\end{lstlisting}

Unlike the original versions of methods \textit{(1)} and \textit{(2)}, the curried versions both take a single argument of the same type; only the types of their return values differ (\textids{System.Int32 -> System.Int32} and \textids{System.Int32}, respectively). This kind of method overloading violates restrictions imposed by the .NET type system.

Favalon, on the other hand, allows multiple function definitions that only differ in their return types. Overload resolution first checks the types of each parameter in the order they are declared before finally checking the type of the return value. If there is any ambiguity during overload selection, a type annotation can be added to the offending item as shown in the following attempt to bind a function:

\begin{lstlisting}[language=FSharp]
  // 'result' is an ambiguous type:
  // (System.Int32 -> System.Int32) |
  // System.Int32
  result = System.Convert.ToInt32 "123"

  // method (2) is selected with the type
  // annotation
  result: System.Int32 =
    System.Convert.ToInt32 "123"
\end{lstlisting}

Because Favalon was designed to behave like an interactive shell does, it prioritizes overloads that return literals in a REPL or any other environment that expects to receive those values. This feature allows Favalon to be used like an ordinary shell despite its functional programming roots:

\begin{lstlisting}[language=FSharp]
  // System.Int32 does not require a function
  // application, so it is automatically
  // selected as the result in the REPL
  System.Convert.ToInt32 "123"
  123
\end{lstlisting}

If more than one overload returns a literal, each candidate is ranked according to the type of its return value as shown in \tabref{tab:overload}. If all of the candidates have the exact same priority, the first one is chosen unpredictably. For example, \textids{System.IFormattable} and \textids{System.IComparable} are both reference types and neither can be chosen on the basis of whether it has an implicit narrowing conversion. As a result, the selected type is indeterminate.

\begin{table*}[bt]
  \caption{Overload Resolution Priority}
  \label{tab:overload}
  \centering
  \begin{tabular}{cl}
    \toprule
    Priority&Types\footnotemark[5]\\
    \midrule
    1 & \textid{System.Byte}, \textid{System.Int16}, \textid{System.Int32}, \textid{System.Int64}\\
    2 & \textid{System.SByte}, \textid{System.UInt16}, \textid{System.UInt32}, \textid{System.UInt64}\\
    3 & \textid{System.Boolean}, \textid{System.Char}, \textid{System.Single}, \textid{System.Double}\\
    4 & String type (\textid{System.String})\\
    5 & Enum types (values that inherit from \textid{System.Enum})\\
    6 & Value types (values that inherit from \textid{System.ValueType})\\
    7 & Class reference types\footnotemark[6]\\
    8 & Interface reference types\footnotemark[6]\\
    \bottomrule
  \end{tabular}
\end{table*}

\footnotetext[5]{Types are listed in order of descending priority on each row.}
\footnotetext[6]{If implicit type conversions are available for both input types, narrowing conversions take priority over widening ones.}

It remains to be seen whether users will actually have difficulty understanding these variations when candidates are selected unpredictably. It's reasonable to assume that users enter expressions into a shell with an understanding of what they're trying to do, so the chances of getting an unexpected result are probably low. If all else fails, the user can always add type annotations to guide Favalon toward a determinate result.

\subsubsection{Literals}

Strings, numeric values, and other .NET literals are expressed as instances of \textids{ConstantTerm}.
Since the type on the .NET runtime has already been identified for the literals. The higher-kinded type can be identified without inference.

\subsubsection{Identifiers}

\textids{VariableTerm} is an identifier. Though its only purpose is to distinguish between IDs, it can be converted into a concrete \textids{Expression} instance by inquiries to Favalon's type system during $\beta$-reduction.

Each sequence of characters that is converted into an \textids{Identity} by the lexer is treated equally. Favalon uses many words and symbols that are similar to reserved words, so its syntax can be freely extended through binding.

The following are some examples of identifiers:

\paragraph*{Bound names}
  \begin{itemize}
    \item Identifiers used in a binding expression. Also called a bound variable. For example, given \textids{foo = 123}, \textids{foo} is an identifier.
    \item Internal identifiers such as \textids{bool}, \textids{true}, and \textids{false} defined by Favalon.
  \end{itemize}

\paragraph*{Symbols}
  \begin{itemize}
    \item A sequence of one or more characters that the lexer recognizes as a symbol token.
    \item The bind operator \textids{=} and the lambda arrow operator \textids{->}.
    \item Symbols for arithmetic, algebraic, and logical operations.
  \end{itemize}

\paragraph*{Symbol names defined by .NET libraries}
  \begin{itemize}
    \item Namespaces like \textids{System} and \textids{System.Collections.Generic}.
    \item Type names like \textids{Int32} and \textids{List}.
    \item Member names like \textids{ToInt32} and \textids{Add}.
    \item Special symbol names like \textids{op_Addition}, which is handled like a plus sign (these symbol names are specified by .NET metadata rules~\cite{ecma335opoverload}).
  \end{itemize}

An higher-kinded expression can be defined using \textids{VariableTerm}. The entire type inference process described here (beginning in \S\ref{sec:inference}) is applied recursively so that even ID types are converted from the type environment into concrete expressions. The following example shows how a \textids{VariableTerm} representing a literal symbol would be looked up in the type environment and replaced with another expression recursivily.

\begin{lstlisting}[language=CSharp,morekeywords={Variable,Type,*}]
  // Before
  Variable(
    foo,
    Variable(System.Int32))

  // After(Inferred TypeTerm)
  Variable(
    foo,
    Type(System.Int32, *))
\end{lstlisting}

\textids{TypeTerm} represents a .NET runtime type expression, and a higher-kinded expression ('*') represents a corresponding higher-kinded type \textids{VariableTerm(*)}.

\subsubsection{Runtime Type Identifiers}

The following code snippets show how .NET type names can be written as identifiers in a shell script:

\begin{lstlisting}[language=FSharp]
  // Integer
  System.Int32

  // ArrayList class
  System.Collections.ArrayList

  // Function application (List<System.String>)
  // on an open generic list (List<>)
  System.Collections.Generic.List System.String
\end{lstlisting}

.NET supports generics; when a generic type is not given any type arguments, it is called an \textit{open generic type}. If these symbols are understood to be identifiers, they can be interpreted as follows:

\begin{lstlisting}[language=FSharp,morekeywords={*}]]
  // Representation of a type itself:
  //   System.Collections.Generic.List<>
  System.Collections.Generic.List: *

  // Type constructor:
  //   System.Collections.Generic.List<T>
  System.Collections.Generic.List: * -> *

  // Value constructor:
  //   System.Collections.Generic.List<
  //     System.String>(System.Int32 capacity)
  System.Collections.Generic.List:
    System.Int32 ->
      System.Collections.Generic.List
      System.String
\end{lstlisting}

These are all considered to be overloaded functions.

Any term that represents a type itself is defined using \textids{TypeTerm}. Any type constructor, which behaves like a type-level function, is represented using \textids{TypeConstructorTerm}. Value constructors, like methods, are treated as functions.

When multiple arguments are given for a generic type, it is assembled and resolved using the same type inference algorithm described earlier for method overloading at the type level. The following example shows how Favalon interprets .NET generic tuple types:

\begin{lstlisting}[language=FSharp]
  // System.Tuple<T1, T2>
  System.Tuple:
    T1 -> T2 -> System.Tuple T1 T2

  // System.Tuple<T1, T2, T3>
  System.Tuple:
    T1 -> T2 -> T3 -> System.Tuple T1 T2 T3

  // System.Tuple<T1, T2, T3, T4>
  System.Tuple:
    T1 -> T2 -> T3 -> T4 ->
      System.Tuple T1 T2 T3 T4
\end{lstlisting}

\subsubsection{Runtime Method Type Information}

\textids{MethodTerm} represents a type expression for a method on the .NET runtime. For example:

\filbreak

\begin{lstlisting}[language=FSharp]
  // System.Int64 System.Math.BigMul(
  //   System.Int32 a, System.Int32 b)
  System.Math.BigMul:
    System.Int32 -> System.Int32 -> System.Int64

  // System.Int32 System.Math.Max(
  //   System.Int32 a, System.Int32 b)
  System.Math.Max:
    System.Int32 -> System.Int32

  // System.Int64 System.Math.Max(
  //  System.Int64 a, System.Int64 b)
  System.Math.Max:
    System.Int64 -> System.Int64
\end{lstlisting}

Method overloading uses different types---like the \textids{Max} function here---and is thus handled the same way as constructor overload resolution. Turning our attention away from static methods for a moment, we can write type expressions for instance methods as follows:

\begin{lstlisting}[language=FSharp]
  // System.Int32 System.Collections.ArrayList.
  //   IndexOf(System.Object item)
  System.Collections.ArrayList.IndexOf:
    System.Object -> System.Int32 ->
      System.Collections.ArrayList
\end{lstlisting}

Instance methods are considered to be functions with an additional parameter that takes a reference to the current object instance (equivalent to \textids{this}).
Define to specify the arguments in curried form, and specify the instance as the last argument.

The \textids{IndexOf} method must be specified using its fully qualified name (including its namespace) as shown here. This limitation is also mentioned in \S\ref{sec:conclusion}.

\subsection{Bound Attributes}

During its lexical analysis and parsing phases, a typical language processor pays attention to the following attributes as it builds a syntax tree according to the types of the ID tokens it encounters:

\begin{itemize}
  \item Prefix vs. infix
  \item Left vs. right associativity
  \item (Operator) precedence
\end{itemize}

In most cases, these attributes are applied to tokens that have been recognized as operators.

Favalon is a bit different---it does not make any decisions based on these attributes during lexical analysis or parsing because doing so would cause those rules to restrict the language's syntax. In a language like C, for example, arithmetic and other operators' attributes, associativity, and precedence are fixed and cannot be redefined.

Favalon instead checks these attributes and modifies the syntax tree at the same time that it performs type inference and $\beta$-reduction. Because operators are represented as ID tokens, these attributes can be applied to them along with any other ID. For example, the user can redefine Favalon's syntax according to whatever script it will be used with:

\filbreak

\begin{lstlisting}[language=CSharp,morekeywords={&&,and}]
  // C-style logical AND
  // (the && operator uses infix notation)
  a && b

  // BASIC-style logical AND
  // (the 'and' token uses infix notation)
  a and b
\end{lstlisting}

Favalon's ID tokens are associated with bound names and used to define nominal types. Though the type system takes care of associating tokens with bound concrete expressions, this attribute data can still be fetched along with any ID's associated expression. The following pseudocode for a binding expression demonstrates this concept:

\begin{lstlisting}[language=FSharp]
  // Arrow operator definition
  // (infix, right-associative)
  let (-> @ INFIX,RTL) = fun param -> fun body -> ...
\end{lstlisting}

\subsubsection{Prefix and Infix Attributes}

Function applications are generally written from left-to-right as a function name followed by an argument. Infix operators are declared in a different order, but once they are converted into prefix notation they can be used just like an ordinary function application.

\begin{lstlisting}[language=FSharp,morekeywords={+}]]
  // Ordinary '+' operator notation (infix)
  123 + 456

  // We can convert the '+' operator into
  // prefix notation so it can be viewed as
  // a function call
  // +: System.Int32 -> System.Int32 ->
  //      System.Int32
  + 123 456
\end{lstlisting}

Favalon's parser considers all tokens to be left-associative, so infix operators must be transformed into prefix notation as shown in \figref{fig:preinfix}.

\begin{figure}[h]
\begin{lstlisting}[language=CSharp,morekeywords={Apply,+}]
  // Pattern #1

  // Syntax tree for '123 + 456' after
  // left-associative function application
  Apply(
    Apply(
      123,
      +),
    456)

  // Syntax tree for '123 + 456' converted to
  // prefix notation
  Apply(
    Apply(
      +,
      123),
    456)

  // Pattern #2

  // Syntax tree for 'abc 123 + 456' after
  // left-associative function application
  Apply(
    Apply(
      Apply(
        abc,
        123),
      +),
    456)

  // Syntax tree for 'abc 123 + 456' converted to
  // prefix notation
  Apply(
    Apply(
      Apply(
        abc,
        +),
      123),
    456)
\end{lstlisting}
\vspace{-5pt}  
\caption{Transforming infix operators into prefix notation}
\label{fig:preinfix}
\end{figure}

\subsubsection{Left and Right Associativity}

Function application is always left-associative. However, some operators may be right-associative, so Favalon transforms on the syntax tree for any ID tokens that correspond to those operators, as shown in \figref{fig:associativity}.

In this case, the target of the right-associative to the argument part of the function application before the transformation.
In the \textids{x -> y} expression, the \textids{y} expression corresponds to the argument part, and becomes the right-associative.

\begin{figure}[h]
\begin{lstlisting}[language=CSharp,morekeywords={Apply,->}]
  // Pattern #1

  // Syntax tree for 'a -> b c' after
  // left-associative function application
  Apply(
    Apply(
      Apply(
        a,
        ->),
      b),
    c),

  // The arrow operator uses infix notation and
  // is right-associative, so the syntax tree
  // must be transformed as follows:
  // -> a (b c)
  Apply(
    Apply(
      ->,
      a),
    Apply(
      b,
      c))

  // Pattern #2

  // Syntax tree for 'a b -> c d' after
  // left-associative function application
  Apply(
    Apply(
      Apply(
        Apply(
          a,
          b),
        ->),
      c),
    d)

  // The arrow operator uses infix notation and
  // is right-associative, so the syntax tree
  // must be transformed as follows:
  // ((a ->) b) (c d)
  Apply(
    Apply(
      Apply(
        a,
        ->),
      b),
      Apply(
        c,
        d)))
\end{lstlisting}
\vspace{-5pt}  
\caption{Performing a right rotation on the syntax tree}
\label{fig:associativity}
\end{figure}

\subsection{Operator Overloading}

Using the core engine features described in the previous sections, we will now define the operators necessary for basic computation with the lambda calculus.

\subsubsection{Arrow Operator}

The arrow \textids{->} is the core operator of the lambda calculus and is required to define functions in a shell script. \figref{fig:arrow} shows how the arrow operator is defined as a syntax tree.

\begin{figure}[h]
\begin{lstlisting}[language=CSharp,morekeywords={Apply,Bind,Lambda}]
  // let (-> @ INFIX,RTL) =
  //   fun param -> fun body -> Reduce(param, body)
  Bind(
    ->,
    INFIX | RTL,
    Lambda(
      param,
      Lambda(
        body,
        // Reduce(): Low level reduce implementation:
        // 1. Store 'param' identity into type environment.
        // 2. Reduce 'body' with type environment.
      )))
\end{lstlisting}
\vspace{-5pt}  
\caption{Syntax tree for the arrow operator}
\label{fig:arrow}
\end{figure}

The innermost \textids{LambdaExpression} uses expression-specific code that cannot be implemented in a Favalon shell script; a C\# implementation was tested instead.

\subsubsection{Arithmetic and Logical Operators}

The special symbol names defined by .NET libraries for arithmetic and logical operations (e.g. \textids{op_Addition}) are automatically recognized and defined as operator overloads. However, syntax trees must be defined separately for these operations because primitive types like \textids{System.Int32} are directly converted into Common Intermediate Language~(CIL) by the C\# and F\# compilers and do not have any equivalent symbol names. \figref{fig:plus} gives an example of a syntax tree for the addition operator with \textids{System.Int32} values.

\begin{figure}[h]
\begin{lstlisting}[language=CSharp,morekeywords={Bind,Lambda}]
  // let (+ @ INFIX,LTR) =
  //   fun a -> fun b -> AddImpl(a, b)
  Bind(
    +,
    INFIX | LTR,
    Lambda(
      a,
      Lambda(
        b,
        // AddImpl(): add a with b
      )))
\end{lstlisting}
\vspace{-5pt}  
\caption{Syntax tree for the addition operator}
\label{fig:plus}
\end{figure}

These fundamental operators also use expression-specific code.

\subsubsection{Pipeline Operator}

Almost every programming language natively supports the operators described thus far. Now let's consider the pipeline operator, which is commonly used in shell scripts. Favalon's pipeline operator is based on the same idea as the one in F\# and was originally introduced by Isabelle/ML~\cite{ppsymbol}. The basic premise is to change the order of function application.

As shown in \figref{fig:pipe}, the pipeline operator can be defined with Favalon's binding expressions alone and does not require any expression-specific code.

\begin{figure}[h]
\begin{lstlisting}[language=CSharp,morekeywords={Apply,Bind,Lambda}]
  // let (| @ INFIX,LTR) =
  //   fun f -> fun g -> g f
  Bind(
    |,
    INFIX | LTR,
    Lambda(
      f,
      Lambda(
        g,
        Apply(
          g,
          f))))
\end{lstlisting}
\vspace{-5pt}  
\caption{Syntax tree for the pipeline operator}
\label{fig:pipe}
\end{figure}

Once this operator has been defined, given a natural shell script expression:

\begin{lstlisting}[language=Favalon,morekeywords={echo,wc,|}]
  echo "abc def ghi" | wc
\end{lstlisting}

We can apply a transformation and perform $\beta$-reduction as follows:

\begin{lstlisting}[language=Favalon,morekeywords={echo,wc,|}]
  # Step 1: Transform into prefix notation
  | (echo "abc def ghi") wc

  # Step 2: Given 'fun f -> fun g -> g f',
  #   apply 'f' and 'g' in reverse order
  wc (echo "abc def ghi")
\end{lstlisting}

In this case we can consider the \textids{wc} command to be a function of type \textids{T1 -> T2}. The command's argument represents the standard input stream and is given .NET's \textids{System.IO.Stream} type. If the \textids{echo} command's return value is similarly assumed to be the standard output stream and is given the \textids{System.IO.Stream} type, both types match and the commands are joined by the pipeline.

If we were to replace the \textids{wc} command with a .NET method that accepts a \textids{System.IO.Stream} argument, the pipeline operator would seamlessly join the external \textids{echo} command with the .NET method. In the following script, the \textids{cat} command loads a file and passes it to the \textids{XElement} class's parser.

\filbreak

\begin{lstlisting}[language=Favalon,morekeywords={cat,|}]
  # System.Xml.Linq.XElement
  #   System.Xml.Linq.XElement.Load(
  #     System.Stream stream)
  cat "sample.xml" |
    System.Xml.Linq.XElement.Load

  # Step 1: Transform into prefix notation
  | (cat "sample.xml")
    System.Xml.Linq.XElement.Load

  # Step 2: Given 'fun f -> fun g -> g f',
  #   apply 'f' and 'g' in reverse order
  System.Xml.Linq.XElement.Load
    (cat "sample.xml")
\end{lstlisting}

At a high level, we can see that the output from the \textids{cat} command is passed directly as an argument to the \textids{Load} method in Step 2, seamlessly connecting the external command with the .NET method. The next section gives more information on type definitions for external commands.

\section{Interactive Shell}

So far, we've seen how Favalon handles types. Now we'll use this type system to add types to external commands.

\subsection{External Command Interfaces}

External command interfaces are made up of the following elements.

\begin{description}
  \item [Standard input (stdin)]
    Generally assumed to be text data, but can also be binary data (byte streams). This data doesn't have any particular semantic significance, so Favalon figures out what kind of data the user is being prompted to enter before sending it along.

  \item [Standard output (stdout) \& error (stderr)]
    Output data streams (the opposite of stdin) without any particular rules for how to handle data (like stdin). Stdout and stderr are treated as entirely separate output streams.

  \item [Process exit code]
    An integer value returned by an external command when its process exits.

  \item [Optional parameters]
    Any number of optional parameters that can be specified by the user.
\end{description}

Favalon adds types to the standard input and output streams as described in the following section.

\subsection{Associated Types for Standard Input and Output}

Standard input contains either text or binary data. Input text can be treated as a single long string (\textids{System.String}) or even associated with a slightly more descriptive type like one of the following.

\begin{description}
  \item [\textids{\small System.IO.TextReader}]
    .NET abstract class that can read characters and strings.

  \item [\textids{\small System.IO.Stream}]
    .NET abstract class that can read and write binary data.

  \item [\textids{\small System.Collections.Generic.IEnumerable<System.String>}]
    String iterator that can read a multiline string as a sequence.
\end{description}

The following types are appropriate for output.

\begin{description}
  \item [\textids{\small System.IO.TextWriter}]
    .NET abstract class that can write characters and strings.
  \item [\textids{\small System.IO.Stream}]
    .NET abstract class that can read and write binary data.
  \item [\textids{\small System.Collections.Generic.IEnumerable<System.String>}]
    String iterator that can output a multiline string as a sequence.
\end{description}

However, there isn't necessarily a one-to-one correspondence between these types:

\begin{itemize}
  \item The \textids{System.IO.TextReader} and \textids{System.IO.TextWriter} types fulfill opposing roles but are not assignment compatible with each other.

  \item The same \textids{System.IO.Stream} type is used for both input and output and is thus assignment compatible, but it would not be appropriate to directly assign an input \textids{System.IO.Stream} to an output \textids{System.IO.Stream} (or vice versa) because these streams have a producer-consumer relationship.
\end{itemize}

For this reason, when we identify an external command we should add type conversions around the resulting function definition. The following sample code emulates these additional type conversions, where \textids{echo} is a method defined in a .NET library, \textids{wc} is an external command provided as input, and \textids{wc_raw} is the actual external command.

\begin{lstlisting}[language=Favalon,morekeywords={echo,wc,tws,wc_raw,|}]
  # Input Favalon script
  echo "abc def ghi" | wc

  # With type conversions
  echo "abc def ghi" | tws | wc_raw
\end{lstlisting}

\begin{lstlisting}[language=Favalon,morekeywords={echo,wc,tws,wc_raw,|}]
  # 'echo' method defined
  # in a .NET library
  echo: System.IO.Stream -> System.IO.TextWriter

  # Function that calls
  # the external command 'wc'
  wc_raw: System.IO.Stream -> System.IO.Stream

  # Type conversion function
  tws: System.IO.TextWriter -> System.IO.Stream

  # Equivalent definition for wc:
  # System.IO.TextWriter -> System.IO.Stream
  let wc = fun tw -> wc_raw (tws tw)
\end{lstlisting}

If we only define the type conversion functions that we need and then apply them to bound functions that we define as overloads when importing external commands, we can automatically select overloads based on the types used before and after external commands.

However, unintended commands could be run if overload selection is unpredictable. To address this problem, we can select an overload according to a type priority lookup table (as described in \S\ref{sec:overloading}) that only contains priorities for these conversion functions; types that are most likely to be able to save data have the highest priority. \tabref{tab:ioconv} gives one example of what this might look like.

\begin{table*}[bt]
  \caption{Overload Resolution Priority for Conversion Functions}
  \label{tab:ioconv}
  \centering
  \begin{tabular}{cl}
    \toprule
    Priority&Conversion Function\\
    \midrule
    1 & \small \textid{System.IO.Stream -> System.IO.Stream}\\
    2 & \small \textid{System.IO.TextWriter -> System.IO.Stream}\\
    3 & \small \textid{System.Collections.Generic.IEnumerable System.String -> System.IO.Stream}\\
    \bottomrule
  \end{tabular}
\end{table*}

\subsubsection{More Expressive Type Definitions}
\label{sec:typedef}

We can use automatic type conversion functions like the ones described in the previous section to achieve stricter typing based on the given data format. For example, consider the case in which a CSV file is provided as input:

\begin{lstlisting}[language=Favalon,morekeywords={cat,|}]
  # Get distinct values from
  # the first column of the CSV file
  cat sample.csv                           |
      (System.Linq.Enumerable.ElementAt 0) |
      System.Linq.Enumerable.Distinct
\end{lstlisting}

It would be more convenient for the \textids{cat} command in this Favalon script to output a \textids{System.String[]} sequence. In other words, we would like to parse the CSV file and convert each row into an array of strings (one for each column) that are then sent one row at a time to \textids{ElementAt}.

If we were to define the \textids{cat} command using the types mentioned in the previous section, its output would either be handled as a byte stream with \textids{System.IO.Stream} or read one line at a time as strings with \textids{System.IO.TextReader} rather than being split into an array of substrings for each column using commas as delimiters. However, we can define a function that will do the conversion for us:

\begin{lstlisting}[language=Favalon,morekeywords={cat,pcsv,|}]
  # Conversion function for parsing CSV data
  pcsv: System.IO.TextReader ->
        System.Collections.Generic.IEnumerable System.String[]
\end{lstlisting}

Now it's trivial to combine the commands in a Favalon script:

\begin{lstlisting}[language=Favalon,morekeywords={cat,pcsv,|}]
  # Use the CSV conversion function
  cat sample.csv                           |
      pcsv                                 |
      (System.Linq.Enumerable.ElementAt 0) |
      System.Linq.Enumerable.Distinct
\end{lstlisting}

\section{Conclusion and Future Work}
\label{sec:conclusion}

In this paper, we have explored techniques for building and structuring an interactive shell based on a typed lambda calculus.

We saw how Favalon performs lexical and syntactic analysis on a script using a very simple set of rules; semantically annotates each token using bound attributes; and redefines shell script syntax as a series of function calls. These features allow external commands to be handled just like existing library methods.

We also saw how pipelines, a fundamental feature of shell scripts, can be viewed as function applications. We examined how Favalon is able to add types to standard input and output for external commands and use those types to examine expressions. By adding overload resolution to the selection of bound functions, Favalon can even provide type inference capabilities while maintaining the convenience of conventional shell script syntax.

There are still several areas of inquiry that have yet to be studied. In the following sections, we will review what remains to be done as future work.

\subsection{Operator Precedence}

A precedence is usually associated with each operator in an expression, but the author has not studied techniques for implementing operator precedence in Favalon nor how practical it would be. One possible approach could involve specifying operator precedence as a bound attribute.

\subsection{Additional Operators}

Definitions for the following items would allow a variety of operations to be written entirely in a Favalon script:

\begin{itemize}
  \item Pattern matching expressions
  \item \textids{if}, \textids{while}, and other flow control statements
  \item Operators for algebraic data types
\end{itemize}

Although any term can be implemented in C\# or F\#, letting users write functions like \textids{pcsv} (as defined in \S\ref{sec:typedef}) themselves would make Favalon an even more useful shell.

\subsection{Namespaces and Higher-Kinded Types}

.NET type definitions include namespaces. Because these are delimited by periods, we could treat the period as a kind of operator representing a function application on each namespace symbol.

When structured in this way, a namespace specification could be handled as a Favalon expression and thus contribute to the overall consistency of the language syntax. In this case, the types of the symbol names would not be defined by .NET, so it may be possible to define them using a namespace-specific higher-kinded type.

\subsection{Instance Member Accessors}

Favalon does not currently account for instance member access. When an instance member is referenced in C\# or F\#, the instance and member names are separated by a period. We could use a period operator for member access just as we could use it on namespaces (as described earlier), allowing the operation to be reduced to a function application.

\subsection{Directory and File Paths}

In order to be used for everyday tasks, an interactive shell must provide a natural syntax for specifying directory and file paths. The simplest way to handle paths is to treat them as simple strings, but shells typically don't require every path to be surrounded by quotes and instead handle them using some kind of ID token. Paths can be thought of as a type of namespace, so by treating path delimiters as operators they too can be reduced to function applications.

Namespaces and instance members both use a period as a delimiter; this is unlikely to be mistakenly processed by the lexer anywhere other than within a decimal number. The situation is a bit more complicated with file paths:

\begin{itemize}
  \item In a POSIX-compliant environment, a forward slash may be misinterpreted as the prefix of an optional argument.
  \item On Windows, a backslash may be misinterpreted as an escape character in a string.
\end{itemize}

Further study is needed to determine what, if any, problems there are in practice.

\subsection{Package System Integration}

The .NET ecosystem encompasses the NuGet package manager for .NET library distribution as well as PowerShell modules (primarily PowerShell-specific functions and cmdlets). If Favalon's syntax allowed these to be used directly, Favalon could be applied to a variety of different tasks and thus leave room to consider other techniques and implementations.

\subsection{Source Code References}

Using complete or partial type inference results, Favalon could publish an API for referencing locations in source code that could then be used to implement command-line completion. With a Language Server Protocol~\cite{lsp} implementation, even text editors could access Favalon's type inference results to provide supplemental functionality.

\subsection{Automatic Definitions for Optional Parameters}

Each external command takes a different number of optional parameters. OCommand's type information was added through a hand-coded DSL, but there is room to consider how this process could be partially automated. For example:

\begin{itemize}
  \item There are some ways of providing information related to command-line arguments in a common format, such as the IEEE POSIX utility argument syntax~\cite{posixuas} and GNU getopt~\cite{gnugetopt}. If we could come up with an algorithm for analyzing help messages (which can often be displayed with the \textids{--help} option), we may be able to acquire information on optional parameters and automatically add types to at least some of them.

  \item .NET binaries can be analyzed using reflection. As a result, it may be possible to easily acquire the information we need by targeting frequently-used libraries for analyzing command-line arguments.
\end{itemize}

\subsection{Compilation}

Even though Favalon runs on the .NET runtime, its syntax trees are run in an interpreter via $\beta$-reduction. The choice to use an interpreter was motivated by a desire to keep the language compact and allow it to be embedded in other applications, but it should also be possible to generate a .NET binary while interpreting the syntax tree. We leave this as a topic for future work.

\begin{acknowledgment}
To Yosuke Morimoto, for all his advice and help reviewing this paper.
\end{acknowledgment}

\begin{flushleft}
  \bibliography{Favalon-en}

\begin{thebibliography}{10}

\bibitem{posixshell}
IEEE: {\em Shell Command Language}, chapter~2 (online),
  \urle{https://pubs.opengroup.org/onlinepubs/9699919799/}.
\newblock  \cite{posix} (2018).

\bibitem{goexec}
Griesemer, R., Pike, R. and Thompson, K.: Package exec runs external commands.,
  Google Inc. (online), \urle{https://golang.org/pkg/os/exec/}
  \refdatee{2020-02}.

\bibitem{gore}
Otsubo, H.: Yet another Go REPL that works nicely. Featured with line editing,
  code completion, and more., Hironao Otsubo (online),
  \urle{https://github.com/motemen/gore/} \refdatee{2020-02}.

\bibitem{fsinteractive}
Carter, P.: Interactive Programming with F\#, Microsoft corporation (online),
  \urle{https://docs.microsoft.com/en-us/dotnet/fsharp/tutorials/fsharp-interactive/}
  \refdatee{2020-02}.

\bibitem{jupyter}
Perez, F. and Granger, B.~E.: Project Jupyter: Computational Narratives as the
  Engine of Collaborative Data Science, Project Jupyter (online),
  \urle{http://archive.ipython.org/JupyterGrantNarrative-2015.pdf}
  \refdatee{2020-02}.

\bibitem{ipython}
Perez, F. and Granger, B.~E.: IPython: A System for Interactive Scientific
  Computing, {\em IEEE Computing in Science \& Engineering},  Vol.~9, No.~3,
  pp.\ 21--29 (online), \doi{10.1109/MCSE.2007.53} (2007).

\bibitem{monadpsh}
Snover, J.~P.: Monad Manifesto, Microsoft corporation (online),
  \urle{https://devblogs.microsoft.com/powershell/monad-manifesto-the-origin-of-windows-powershell/}
  \refdatee{2020-02}.

\bibitem{psise}
Microsoft: The Windows PowerShell ISE, Microsoft corporation (online),
  \urle{https://docs.microsoft.com/en-us/powershell/scripting/components/ise/introducing-the-windows-powershell-ise?view=powershell-7}
  \refdatee{2020-02}.

\bibitem{shcaml}
Heller, A. and Tov, J.~A.: Caml-Shcaml: An OCaml Library for Unix Shell
  Programming, {\em ACM SIGPLAN workshop on ML},  Vol.~8, pp.\ 79--90 (online),
  \doi{10.1145/1411304.1411316} (2008).

\bibitem{ocommand}
Asakura, I., Masuhara, H. and Aotani, T.: OCommand: A Domain Specific Language
  for Type Safe Shell Programming in OCaml, {\em Information Processing Society
  of Japan. Transactions on programming},  Vol.~7, No.~5, pp.\ 14--14 (online),
  \urle{http://prg.is.titech.ac.jp/wp-content/uploads/2014/08/paper.pdf}
  (2014).

\bibitem{tcsh}
Greer, K.: C shell with command and filename recognition/completion, Ken Greer
  (online),
  \urle{https://groups.google.com/forum/?hl=en#!msg/net.sources/BC0V7oosT8k/MKNdzEG_c3AJ}
  \refdatee{2020-02}.

\bibitem{bashac}
GNU: Programmable Completion, GNU Free Documentation, Free Software Foundation
  (online),
  \urle{https://www.gnu.org/software/bash/manual/html_node/Programmable-Completion.html}
  \refdatee{2020-02}.

\bibitem{unicodepunctuationopen}
Unicode: UCA: Punctuation-Open, Unicode Consortium (online),
  \urle{https://www.unicode.org/charts/normalization/chart_Punctuation-Open.html}
  \refdatee{2020-10}.

\bibitem{unicodepunctuationclose}
Unicode: UCA: Punctuation-Close, Unicode Consortium (online),
  \urle{https://www.unicode.org/charts/normalization/chart_Punctuation-Close.html}
  \refdatee{2020-10}.

\bibitem{algorithmw}
Milner, R.: A theory of type polymorphism in programming, {\em Journal of
  Computer and System Sciences},  Vol.~17, No.~3, pp.\ 348--375 (online),
  \doi{10.1016/0022-0000(78)90014-4} (1978).

\bibitem{ecma335overload}
Ecma: {\em Overloading}, chapter\ 1.10.2 (online),
  \urle{https://www.ecma-international.org/publications/standards/Ecma-335.htm}.
\newblock  \cite{ecma335} (2012).

\bibitem{ecma335opoverload}
Ecma: {\em Operator overloading}, chapter\ 1.10.3 (online),
  \urle{https://www.ecma-international.org/publications/standards/Ecma-335.htm}.
\newblock  \cite{ecma335} (2012).

\bibitem{ppsymbol}
Nipkow, T., Paulson, L.~C. and Wenzel, M.: Archeological Semiotics: The Birth
  of the Pipeline Symbol, Microsoft corporation (online),
  \urle{https://docs.microsoft.com/en-us/archive/blogs/dsyme/archeological-semiotics-the-birth-of-the-pipeline-symbol-1994}
  \refdatee{2020-02}.

\bibitem{lsp}
Team, T. V.~C.: A Common Protocol for Languages, Microsoft corporation
  (online),
  \urle{https://code.visualstudio.com/blogs/2016/06/27/common-language-protocol}
  \refdatee{2020-02}.

\bibitem{posixuas}
IEEE: {\em Utility Argument Syntax}, chapter\ 12.1 (online),
  \urle{https://pubs.opengroup.org/onlinepubs/9699919799/}.
\newblock  \cite{posix} (2018).

\bibitem{gnugetopt}
GNU: Parsing Long Options with getopt_long, The GNU C Library, Free Software
  Foundation (online),
  \urle{https://www.gnu.org/software/libc/manual/html_node/Getopt-Long-Options.html}
  \refdatee{2020-02}.

\bibitem{posix}
IEEE: The Open Group Base Specifications Issue 7, 2018 edition, IEEE and The
  Open Group (online), \urle{https://pubs.opengroup.org/onlinepubs/9699919799/}
  \refdatee{2020-02}.

\bibitem{ecma335}
Ecma: Standard ECMA-335, Common Language Infrastructure (CLI) 6th edition, Ecma
  International (online),
  \urle{https://www.ecma-international.org/publications/standards/Ecma-335.htm}
  \refdatee{2020-02}.

\end{thebibliography}
\end{flushleft}

  
\end{document}